\documentclass[%
 reprint,
 superscriptaddress,
 preprintnumbers,
 nofootinbib,
 longbibliography,
 amsmath,
 amssymb,
 aps, 
 showpacs,
prx,
]{revtex4-2}
\pdfoutput=1
\usepackage[utf8]{inputenc}
\usepackage{dsfont}
\usepackage{caption}
\usepackage{subcaption}
\usepackage{orcidlink}
\usepackage{amsmath, amssymb, amscd, amsfonts, mathtools, physics, hyperref, graphicx, color}

\begin{document}

\title{Scaling Monte-Carlo-Based Inference on Antibody and TCR Repertoires}

\author{Josiah Couch\orcidlink{0000-0002-7416-5858}}
 \affiliation{Department of Pathology, Beth Israel Deaconess Medical Center, Boston, MA 02215}

\author{Rohit Arora\orcidlink{0000-0001-7128-6403}}
 \altaffiliation{Current address: Novo Nordisk, Cambridge, MA 02142}
 \affiliation{Department of Pathology, Beth Israel Deaconess Medical Center, Boston, MA 02215}

\author{Jasper Braun\orcidlink{0000-0003-1250-4399}}
 \affiliation{Department of Pathology, Beth Israel Deaconess Medical Center, Boston, MA 02215}

\author{Joesph Kaplinsky\orcidlink{0000-0003-3001-4721}}
 \altaffiliation{Current address: Genomics England, One Canada Square, London, E14 5AB, UK }
 \affiliation{Department of Pathology, Beth Israel Deaconess Medical Center, Boston, MA 02215}

\author{Elliot Hill\orcidlink{0009-0004-1987-3749}}
 \altaffiliation{Current address: Department of Biostatistics and Bioinformatics, Duke University School of Medicine, Durham, NC 27710}
 \affiliation{Department of Pathology, Beth Israel Deaconess Medical Center, Boston, MA 02215}

\author{Anthony Li}
 \altaffiliation{Current address: Argenx US Inc. 33 Arch Street, 32nd Floor
 Boston, MA 02110}
 \affiliation{Department of Pathology, Beth Israel Deaconess Medical Center, Boston, MA 02215}

\author{Brett Altschul\orcidlink{0000-0002-3297-7815}}
 \affiliation{Department of Physics and Astronomy, University of South Carolina, Columbia, SC 29208}
 
\author{Ramy Arnaout\orcidlink{0000-0001-6955-9310}}
 \email{rarnaout@bidmc.harvard.edu}
 \affiliation{Department of Pathology, Beth Israel Deaconess Medical Center, Boston, MA 02215}
 \affiliation{Harvard Medical School, Boston, MA 02115}

\date{\today}

\begin{abstract}

    Previously, it has been shown that maximum-entropy models of immune-repertoire sequence can be used to determine a person's vaccination status.  However, this approach has the drawback of requiring a computationally intensive method to compute each model's partition function ($Z$), the normalization constant required for calculating the probability that the model will generate a given sequence. Specifically, the method required generating approximately $10^{10}$ sequences via Monte-Carlo simulations for each model. This is impractical for large numbers of models. 
    Here we propose an alternative method that requires estimating $Z$ this way for only a few models: it then uses these expensive estimates to estimate $Z$ more efficiently for the remaining models.
    %
    %
    We demonstrate that this new method enables the generation of accurate estimates for 27 models using only three expensive estimates, thereby reducing the computational cost by an order of magnitude. Importantly, this gain in efficiency is achieved with only minimal impact on classification accuracy. Thus, this new method enables larger-scale investigations in computational immunology and represents a useful contribution to energy-based modeling more generally.
    
\end{abstract}

\maketitle

\tableofcontents

\section{Introduction}

Energy-based models (EBMs) generally---and maximum entropy (MaxEnt) models particularly---have a wide range of applications, including in statistical physics~\cite{boltzmann1868, gibbs1876, gibbs1878, jaynes1957a, jaynes1957b} (where the major statistical ensembles all take the form of EBMs), natural language processing~\cite{lafferty1995, berger1996}, finance~\cite{molins2004}, RNA~\cite{yeo2004} and protein~\cite{Shimagaki2019} sequence motifs, ecology~\cite{shipley2006, phillips2006, williams2009}, modeling flocking behavior in birds~\cite{cavagna2014}, modeling voting behavior~\cite{lee2015}, describing patterns of activity in neurons~\cite{ferrari2017, Nghiem2018}, modeling disease outbreaks \cite{Ansari2022}, modeling the environmental preferences of plant pathogens~\cite{cohen2023}, and modeling immune repertoires~\cite{mora_maximum_2010, Arora519108}, among many others~\cite{de_martino2018}. As probabilistic models, EBMs can be used as generative models when coupled with Monte Carlo sampling methods. When properly normalized, they can, unlike most types of discriminative models, also be used for Bayesian inference. In nontrivial settings, however, normalizing EBMs (or indeed any probabilistic models based on initially unnormalized probabilities) can be computationally quite expensive.

In previous work, MaxEnt models were trained on antibody heavy-chain (IGH) and T-cell receptor $\beta$-chain (TRB) repertoires' third complementary-determining regions (CDR3s), using features based on the physicochemical properties of their constituent amino acids~\cite{Arora519108}. It was demonstrated that these models allowed for the classification of influenza vaccination status among 31 samples from 14 individuals. However, this classification required the estimation of partition functions: the normalization constants of the individual probability distributions. This was done using in-house Monte-Carlo (MC) based estimation software,
which was computationally expensive. The partition function for a given model is usually abbreviated as $Z$.

To understand the computational difficulty of the problem, consider a model that represents the distribution of amino-acid sequences comprising some collection of proteins, such as TCRs or B-cell receptors (BCRs) (e.g. IGH). Consider specifically the set of all possible polypeptide chains 100 amino acids in length, which is approximately the length of a TRB or IGH variable region. There are $20^{100} \approx 10^{130}$ such sequences. Thus an exact computation of the partition function would involve a sum of $~10^{130}$ terms. Such a sum is infeasible with present-day computational resources, even before considering that we will likely want to normalize many such models (e.g. one per person per timepoint). In a few special instances, such as the one-dimensional and two-dimensional local Ising or Potts models, there are shortcuts to computing this sum. It is very unlikely such simplifications will exist in general, however, as the problem of computing partition functions has been shown to be \#P-hard~\cite{agrawal2021, roth1996}. Of course in practice, we do not need an exact result, and methods such as bridge sampling~\cite{bennett1976, Meng1996, Gronau2017} exist precisely in order to approximate these kinds of sums more efficiently. Yet even in these cases, it may be necessary to generate an enormous Monte Carlo sample from the model in question. We asked whether we could improve the efficiency of estimating $Z$ for each model without sacrificing classification accuracy, using immune repertoires as a test case. 
%


\subsection{Energy-Based Models}

An EBM is a  model that assigns an unnormalized probability $U_{\vec{\theta}}(x)$ to every potential state $x$ (meaning, in the context of CDR3 repertoires, every possible amino acid sequence up to some maximum length) based on a parameterized energy function $E(x, \vec{\theta}\,)$ according to
\begin{equation}
    U_{\vec{\theta}}(x) = e^{- E(x, \vec{\theta}\,)}.
\end{equation}
The models used in this paper are MaxEnt models, in which the energy takes the form
\begin{equation}
    E(x, \vec{\theta}\,) := E_{\vec{\theta}}(x) = \sum_i \theta_i f_i(x)
\end{equation}
for a set of features $\{f_i\}$. Such models were introduced in Refs.~\cite{jaynes1957a, jaynes1957b} and are based on distributions long studied in statistical physics. The name ``maximum entropy'' comes from the fact that these models maximize the entropy of the resulting distribution subject only to constraints on the moments of the features. The parameters $\vec{\theta}$ fix these moments and determine how the distribution is allowed to vary from the uniform distribution (which corresponds to $\vec{\theta} = \vec{0}\,$)~\cite{yeo2004, russ_natural-like_2005, mora_maximum_2010}.

MaxEnt models can be trained using, for example, gradient descent to maximize the likelihood of a training sample as estimated by the model. Gradients of the log likelihood turn out to depend only on the feature moments for the current model, which can be estimated using Monte Carlo methods, and the sample moments of the training sample. 

\subsection{Estimating Partition Functions}

The problem of normalizing an initially unnormalized probability distribution shows up in a number of contexts and has an extensive literature going back several decades. A review of some of this work may be found in section 6 of Ref.~\cite{Neal1993}. In statistical mechanics, such a normalization constant shows up for the various (microcanonical, canonical, etc.) thermodynamic ensembles and is known as the \textit{partition function}\footnote{Strictly speaking, the partition function in statistical mechanics should be understood as the function which, for a parameterized family of distributions, maps parameter values to the corresponding normalization constant, but that is not a distinction we will make here.}, a term we shall use in most of this work, and is usually written $Z$. Knowing the partition function (as a function of the distribution parameters) allows one to compute all the macroscopic physical quantities that characterize the distribution, such as the mean values of the entropy, internal energy, and magnetization, as well as each of their fluctuations. In the context of Bayesian inference, the posterior distribution takes the form of an unnormalized distribution when the distribution of the evidence is unknown. In a few cases (for example, a multitude of models in one spatial dimension, or the Ising model with nearest-neighbor interactions in two dimensions), the partition functions may be computed analytically, but in the typical case an exact solution is intractable. Indeed, the general case has been shown to be \#P-hard~\cite{roth1996}. As a result, approximation schemes---either analytical or computational---typically need to be employed. 

A large class of computational approximation schemes rely on Monte-Carlo methods for generating model samples; these schemes only estimate the ratio of the partition functions of two models. Alternatively, one can view them as estimating the partition function of one model based on the already known partition function of a second model. In the present work, we will refer to these as the \textit{target} and the \textit{teammate}, respectively. Such methods include bridge sampling~\cite{bennett1976} and the free energy perturbation method~\cite{zwanzig1954}---also known as simple importance sampling (SIS)---among others. The method used previously for immune repertoires by Arora et al. \cite{Arora519108} is also in this class. 
Here we focus on estimating the partition functions themselves (though it should be noted that for the task of maximum likelihood inference, strictly speaking all that is needed is the ratio of the partition function of every model to some fixed reference model).

Given two unnormalized probability distributions whose densities are given by
\begin{equation}
    \mathcal{P}_{\vec{\theta}_{0}}(x) \propto e^{-E_{\vec{\theta}_0}(x)}\\
\end{equation}
and
\begin{equation}
    \mathcal{P}_{\vec{\theta}}(x) \propto e^{-E_{\vec{\theta}}(x)}\\
\end{equation}
the normalization constants of these (unnormalized) distributions---i.e. the partition functions of these models---are defined to be
\begin{equation}
    Z(\vec{\theta}\,) = \sum_x e^{-E_{\vec{\theta}}(x)} 
\end{equation}
and
\begin{equation}
    Z(\vec{\theta}_0)  = \sum_x e^{-E_{\vec{\theta}_0}(x)}.
\end{equation}
In the following we consider $\mathcal{P}_{\vec{\theta}}(x)$ as the target distribution and $\mathcal{P}_{\vec{\theta}_{0}}$ as the teammate.

The free energy perturbation method~\cite{geyer1992, Neal1993} estimates $Z(\vec{\theta}\,)$ from $Z(\vec{\theta}_0)$ (or alternatively, estimates their ratio) according to 
\begin{eqnarray}
    \frac{Z(\vec{\theta}\,)}{Z(\vec{\theta}_0)} 
    &=& \frac{ \sum_{x} e^{ - E_{\vec{\theta}}(x)} }{ \sum_{x}  e^{- E_{\vec{\theta}_0}(x)} } \\
    &=& \sum_{x} \frac{ e^{ - E_{\vec{\theta}}(x)} }{  e^{- E_{\vec{\theta}_0}(x)} } \frac{ e^{ - E_{\vec{\theta}_0}(x)} }{ \sum_{x'}  e^{- E_{\vec{\theta}_0}(x')} } \\
    &=&  \expval{  e^{- \left(E_{\vec{\theta}} - E_{\vec{\theta}_0}\right)} }_{\vec{\theta}_0} \label{eq: teamwork}\\
    &\approx& \sum_i e^{- \left[E_{\vec{\theta}}(x_i) - E_{\vec{\theta}_0}(x_i) \right]},
\end{eqnarray}
where $\{x_i\}$ is a Monte Carlo sample drawn from $\mathcal{P}_{\vec{\theta}_0}$. This Monte Carlo procedure will typically provide a good estimate if every region with non-negligible probability under $\mathcal{P}_{\vec{\theta}}$ also has non-negligible probability under $\mathcal{P}_{\vec{\theta}_0}$. Otherwise, it will tend to do poorly~\cite{neal2005}. One way around this is to consider these distributions as part of a set of models $\mathcal{P}_{\vec{\theta}_{\lambda}}$, $\lambda \in [0,1]$, such that the $\vec{\theta}_{\lambda}$ interpolate between $\vec{\theta}_1 := \vec{\theta}$ and $\vec{\theta}_{0}$. One may then estimate~\cite{neal2005}
\begin{eqnarray}
    \frac{ Z(\vec{\theta}\,) }{ Z(\vec{\theta}_0) } &=& \prod_{i=0}^{N} \frac{ Z(\vec{\theta}_{\lambda_{i+1}}) }{ Z(\vec{\theta}_{\lambda_i}) } \\
    &=& \prod_{i=0}^{N} \expval{ 
        e^{ -\left(E_{\vec{\theta}_{\lambda_{i+1}}} - E_{\vec{\theta}_{\lambda_i}}\right) }
    }_{\vec{\theta}_{\lambda_i}},
\end{eqnarray}
with $\lambda_i = \frac{i}{N}$ for some $N>0$.
Bridge sampling was introduced in Ref.~\cite{bennett1976} as the ``acceptance ratio method,'' before being rediscovered in Ref.~\cite{Meng1996}, whose authors coined the term ``bridge sampling''~\cite{neal2005}. Bridge sampling seeks to cure the weaknesses of the free energy perturbation method by using a single intermediate model $\mathcal{P}_{\text{bridge}}(x) \propto e^{-E_{\text{bridge}}(x)}$. One then estimates
\begin{equation}
    \frac{ Z(\vec{\theta}\,) }{ Z(\vec{\theta}_0) }=\frac{ \expval{  e^{- \left(E_{\text{bridge}} - E_{\vec{\theta}_0}\right)} }_{\vec{\theta}_0} }{ \expval{ e^{- \left(E_{\text{bridge}} - E_{\vec{\theta}}\right)} }_{\vec{\theta}} }.
\end{equation}

Both of these methods (and others in this general family) may not perform well (or alternatively may perform well only when the samples used to compute moments are taken to be very large) if the target and teammate distributions are very different (i.e. there is a large distance between them, for an appropriate choice of metric).

In the immune-repertoire example in Ref.~\cite{Arora519108}, 
computing the partition functions required sampling $\geq 10^{10}$ amino acid sequences via Markov-chain Monte Carlo (MCMC) methods. Even on a highly-parallelized high-performance computing cluster, this requires a day or more of running time. This severely limits the practical feasibility of using this technique directly for Bayesian inference, especially in the case where many such models need to be normalized. 
We believe that the main reason this method is so high-cost is that this teammate distribution is very far from the model distribution. In fact, it is essentially a uniformly random distribution at each length, combined with a distribution on lengths that depends on the target. (Immune repertoires include amino-acid sequences of multiple lengths.) The advantage of this distribution is that its normalization factor can easily be calculated exactly. The disadvantage is, it has a much higher entropy than any of the target models. As such, it takes an extremely large sample to encounter most of the states which have a high-probability in the target model. 
In fact, many of the target models are much closer to one-another than they are to their respective teammate models. 

The key observation of this paper is that we can use this proximity to our advantage. Once the first few models have been normalized using the expensive but proven method in \cite{Arora519108}, those few models can be used as alternative teammates for the remaining targets. This strategy can even be used iteratively, with the high-entropy teammates used to normalize the first targets, these targets used as teammates for a second round of targets, this second round of targets used as teammates for a third round, and so on. In the rest of this paper we will show empirically that this method works well, achieving high classification accuracy while significantly speeding up the process of (approximately) normalizing a fairly sizeable batch of IGH and TRB immune-repertoire models.

\section{Methods}\label{sec: methods}

\subsection{Data}

A total of 19 unique repertoires were studied, summarized in table~\ref{fig:data-table}. Seventeen of these were IGH and TRB CDR3 repertoires representing diverse physiological and pathophysiological states, including infection, vaccination, and cancer, as well as repertoires from subjects that lacked these conditions. Two were artificial ``repertoires'' of randomized sequences created from TRB repertoires (see below). The TRB repertoires included three from subjects imputed to be positive for cytomegalovirus (CMV) (``infected''), four imputed to be CMV-negative (``baseline'') \cite{britanova2014}, and two from subjects with breast cancer (``cancer'') \cite{beausang2017}. CMV infection status was imputed as in \cite{arora2022}. The IGH repertoires similarly include two repertoires from subjects who had received an influenza vaccine (``vaccinated'') and four from subjects that had not (``baseline'') \cite{vollmers2013}, as well as two repertoires from subjects with chronic lymphocytic leukemia (``cancer''---although in this case the repertoire includes sequences from the cancerous clone itself, not just sequences elaborated in response to/in the context of the cancer) \cite{bashford-rogers2013}. The random repertoires were created from TRB repertoires \cite{britanova2014} by preserving the lengths of each sequence but randomizing the amino acids among all sequences. As such, these have the same length and single-amino-acid distributions as their source TRB repertoires, but with all correlations between different amino acids, or between amino acids and position in the sequence, randomized away.

\begin{table}
    \begin{tabular}{ | c | c | c | } 
     \hline
     Repertoire Type & Status & Number of Repertoires \\  
     \hline
     TRB & 
     \begin{tabular}{c}
         baseline\\
         infected\\
         cancer
     \end{tabular} &
     \begin{tabular}{c}
         4\\
         3\\
         2
     \end{tabular} \\   
     \hline
     IGH & 
     \begin{tabular}{c}
         baseline\\
         vaccinated\\
         cancer
     \end{tabular} &
     \begin{tabular}{c}
         4\\
         2\\
         2
     \end{tabular} \\ 
     \hline
     Random & - & 2 \\
     \hline
    \end{tabular}
    \caption{Summary of repertoires. For TRB repertoires, \emph{baseline} and \emph{infected} indicate imputed CMV infection status, whereas for IGH repertoires, \emph{baseline} and \emph{vaccinated} indicate influenza vaccination status. Additionally, the TRB cancer repertoires are from breast cancer, whereas the IGH cancer repertoires are from chronic lymphocytic leukemia.}
    \label{fig:data-table}
\end{table}

\subsection{Models}

A total of 29 maximum entropy (MaxEnt) models were trained as described previously \cite{Arora519108}. For reference, each model was named according to its cell type, disease state, feature set, and and a number that along with this other information uniquely identifies the model.
Twenty of these models (one per repertoire, plus a replicate model for one of the random repertoires, as a control) were trained using a set of features consisting of lengths, frequencies of single amino acids in both an entire sequence and in the first and last four amino acids of a sequence (the canonical stems; IGH and TRB proteins adopt stem-loop structures), and sums of pairwise products of physio-chemical descriptors of amino acids between different locations (including nearest neighbors, next-to-nearest neighbors, and opposites (i.e. first with last, second with second from last, etc.)),
and summed over both the entire sequence and just the first and last four amino acids,
as well as products of physio-chemical descriptors of four consecutive amino acids. This was feature set 1. To test a second set of features, nine additional MaxEnt models were trained on a subset of the repertoires: two each on IGH baseline, IGH vaccinated, and TRB baseline repertoires, as well as three on TRB infected repertoires. These were trained using a different 
set of features that did not include products of four physiochemical descriptors, but which did include products between 3rd nearest neighbors. This was feature set 2. Of these 29 models, two fits failed to converge and were thus excluded from the remainder of the study. These were the models trained on the two IGH cancer repertoires, which as described above come from subjects with chronic lymphocytic leukemia. As such, the failure of these models to converge is perhaps unsurprising, given that these repertoires are dominated by a single large clone. Conversely, because these repertoires can be well described by the sequence clone, there is a diminished need for a compact generative model (e.g. a MaxEnt model) to describe them.

\begin{table*}
    \begin{tabular}{ | c | c | c | c | c | c | } 
     \hline
     Name & Repertoire Type & Disease State & Feature Set & Test/Train & Converged  \\  
     \hline
     Random 1-1    & randomers &          - & 1 & train & yes \\ \hline
     Random 1-2    & randomers &          - & 1 & train & yes \\ \hline
     Random 1-3    & randomers &          - & 1 &  test & yes \\ \hline
     TRB b.l. 1-2  & TRB       &   baseline & 1 & train & yes \\ \hline
     TRB b.l. 1-1  & TRB       &   baseline & 1 & train & yes \\ \hline
     TRB b.l. 1-3  & TRB       &   baseline & 1 &  test & yes \\ \hline
     TRB b.l. 1-4  & TRB       &   baseline & 1 &  test & yes \\ \hline
     TRB infex 1-3 & TRB       &   infected & 1 & train & yes \\ \hline
     TRB infex 1-1 & TRB       &   infected & 1 & train & yes \\ \hline
     TRB infex 1-2 & TRB       &   infected & 1 &  test & yes \\ \hline
     TRB can. 1-1  & TRB       &     cancer & 1 & train & yes \\ \hline
     TRB can. 1-2  & TRB       &     cancer & 1 &  test & yes \\ \hline
     TRB b.l. 2-1  & TRB       &   baseline & 2 & train & yes \\ \hline
     TRB b.l. 2-2  & TRB       &   baseline & 2 &  test & yes \\ \hline
     TRB infex 2-3 & TRB       &   infected & 2 & train & yes \\ \hline
     TRB infex 2-1 & TRB       &   infected & 2 & train & yes \\ \hline
     TRB infex 2-2 & TRB       &   infected & 2 &  test & yes \\ \hline
     IGH b.l. 1-1  & IGH       &   baseline & 1 & train & yes \\ \hline
     IGH b.l. 1-4  & IGH       &   baseline & 1 & train & yes \\ \hline
     IGH b.l. 1-2  & IGH       &   baseline & 1 &  test & yes \\ \hline
     IGH b.l. 1-3  & IGH       &   baseline & 1 &  test & yes \\ \hline
     IGH vax 1-2   & IGH       & vaccinated & 1 & train & yes \\ \hline
     IGH vax 1-1   & IGH       & vaccinated & 1 &  test & yes \\ \hline
     IGH b.l. 2-1  & IGH       &   baseline & 2 & train & yes \\ \hline
     IGH b.l. 2-2  & IGH       &   baseline & 2 &  test & yes \\ \hline
     IGH vax 2-1   & IGH       & vaccinated & 2 & train & yes \\ \hline
     IGH vax 2-2   & IGH       & vaccinated & 2 &  test & yes \\ \hline
     IGH can. 1-1  & IGH       &     cancer & 1 & train &  no \\ \hline
     IGH can. 1-2  & IGH       &     cancer & 1 &  test &  no \\ \hline
    \end{tabular}
    \caption{Summary of models. Note that the first and second parts of the name (first only for randomer models) indicates cell type and disease state, the next part indicates the feature set used, and the last number (along with the other information) uniquely identifies that model. }
    \label{fig:model-table}
\end{table*}

\subsection{Partition Function Estimates Using Non-Repertoire Teammates (Previous Method)}

For each model, we first used the previous method~\cite{Arora519108} to estimate the partition function for each model. This method uses the Metropolis-Hastings algorithm to sample from two distributions, the target distribution (one of the immune repertoire models) and a teammate distribution. The teammate distribution was such that the probability of a sequence depends only on its length. 

\begin{figure}
    \centering
    \includegraphics[width=0.48\textwidth]{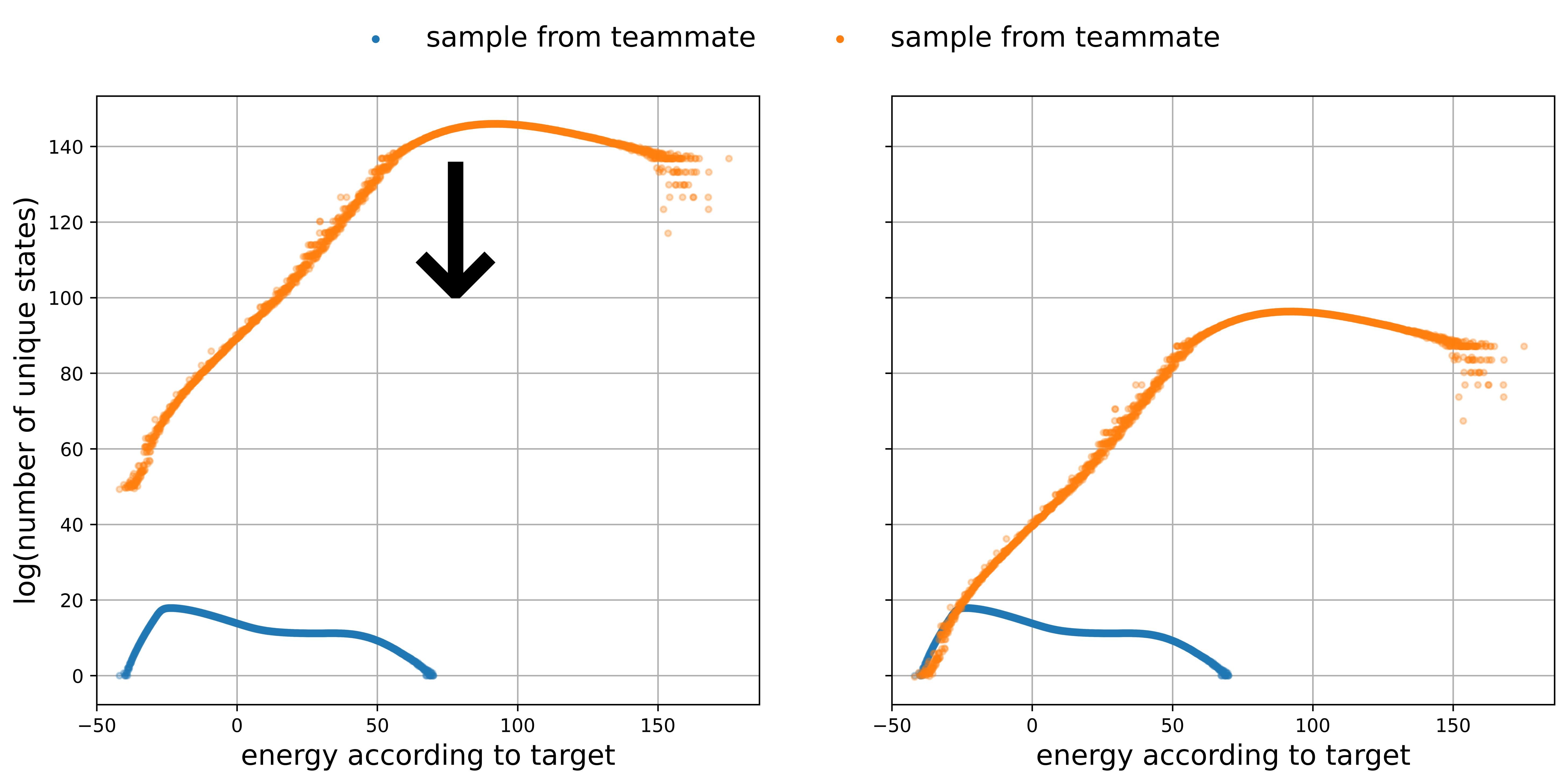}
    \caption{Left: the density of states estimates from both the target and teammate samples for model DCW4o. Right: the same plot, but with the estimate based on the teammate sample rescaled (downward arrow in left panel). The downward shift required to bring the upper distribution into alignment with the lower distribution is the $\ln{Z}$.}
    \label{fig: jabba_curves}
\end{figure}

From each of these samples, we estimated the density of states of the target distribution: the distributions of energies, i.e. (unnormalized) negative log probabilities. We describe the procedure graphically (Fig. \ref{fig: jabba_curves}). For the target sample (green in Fig. \ref{fig: jabba_curves}), this was done by binning the energies and counting the number of unique sequences in each bin. For the teammate distribution (yellow in Fig. \ref{fig: jabba_curves}) each sequence contributed a weight equal to its (unnormalized) probability in the target distribution divided by its (normalized) probability in the teammate distribution. Energies were then binned with the same binning as before, with the weight for each sequence added to the corresponding energy bin. This resulted in two histograms representing the density of states. The first of these was estimated based on a sample of $10^{10}$ Monte Carlo (MC)-generated sequences from the target distribution and represents an absolute estimate; that is, the entries directly estimate the number of unique sequences per bin. The second of these was estimated from a sample of $10^{11}$ MC-generated sequences drawn from the teammate distribution and represents only a relative estimate, in that the overall histogram differs from the (estimated) density of states by an overall multiplicative constant: a vertical shift in Fig. \ref{fig: jabba_curves}.  

We know that scaling the teammate by (the natural logarithm of) the partition function, $\ln{Z}$, will make the absolute probabilities in each bin equal: it will shift the teammate distribution up or down until the high-confidence part of this curve coincides with the high-confidence part of the target distribution. The high-confidence part of the teammate distribution begins when bins contain enough sequences; it will fall off to the left of that (the lowest-energy sequences are unlikely to be sampled with sufficient density by the teammate's random sampling). Meanwhile, the high-confidence part of the target distribution is the leftmost part; the distribution will fall off to the right (higher-energy sequences are unlikely to be sampled sufficiently densely by sampling from the target). The magnitude of the shift required gives the target's $\ln{Z}$. 
Note this method requires a substantial amount of computational effort (typically a day or more on an academic supercomputing cluster) due to the large numbers of sequences sampled. This large sizes are necessary to ensure meaningful overlap between the two density-of-states histograms.

\subsection{Partition Function Estimates with Immune-Repertoire Teammates (New Method)}

Following the bridge-sampling partition function estimates described above, we performed a second analysis using the new method developed for this paper. For each of the 27 models, we computed 26 additional estimates using the free energy perturbation method, one using each of the other 26 models as a teammate. These estimates were computed using a 300,000-sequence sample generated from each model using MCMC methods. These samples were independent of those used to compute the bridge-sampling estimates.


\begin{figure*}
\begin{center}
    \includegraphics[width=0.98\textwidth]{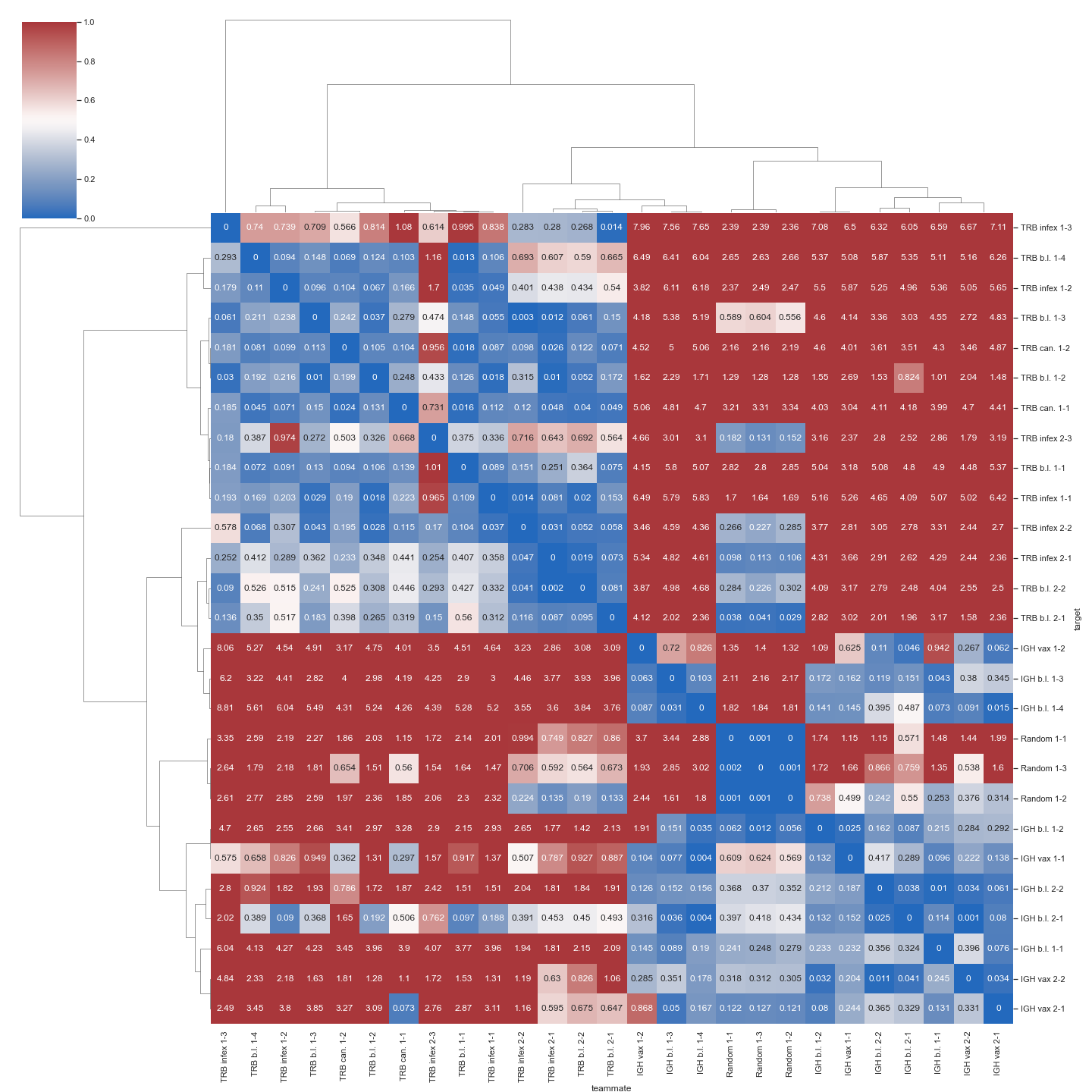}
    \caption{ Differences in log Z estimates using methods (i) and (ii) for each pair (target, teammate) of immune repertoire models.}\label{fig: errors_heatmap_all}
\end{center}
\end{figure*}

For each of these $702$ ($= 27 \text{ targets} \times 26 \text{ teammates})$ free-energy-perturbation estimates, we found an estimated empirical log error, computed as the absolute value of the difference between the estimate in question for the $\ln{Z}$ and that found using the non-immune repertoire teammate. These error estimates ranged from $0.000281$
to $8.81$.
We then trained a model to predict when these errors will fall below a threshold, which we chose, somewhat arbitrarily, to be that there should be less than a 30\% error in the estimated value of $Z$. This threshold translated to empirical log errors of up to $\ln(1.3) \approx 0.262$. Since partition functions for different repertoires are observed to differ by many orders of magnitude, a 30\% error indicates a quite accurate estimate of the relevant $Z$.

 To this end, we divided the 27 models into two sets: a 15-model training set and a 12-model validation set. The details of this split are given in table~\ref{fig:model-table}. Since each individual model is designated as either a training model or a validation model, target-teammate pairs divide naturally into three sets: a training set, where both the target and teammate are from the model training set; a validation set, where both are from the model validation set; and a ``crossover'' set, consisting of the remaining mixed pairs. We trained a random-forest classifier on the training set, achieving a validation accuracy of $89\%$. The input features for this classifier were simple functions of the model parameters: the root-mean-squared difference between the bias vectors for five different types of biases (including two types of first-order bias, two types of second-order bias, and fourth-order biases), as well as a binary variable that flagged when one member of the pair was fit on IGH  but the other was fit on TRB.
%
On the set of all pairs (including validation pairs, training pairs, and crossover pairs), $75\%$ of pairs classified by the model as ``good'' had log errors of $0.215$ or lower ($<24\%$).

The methods described above consider every model as both a target and a teammate, and require previously-computed estimates for the partition functions of all models. However, this was only necessary for the purpose of training the classifier. In the remaining analysis, we simulated the scenario where a seed set of only a small number of models were initially identified as likely good teammates for the remaining models. The previous method outlined above was then used to generate estimates for these few models, and the resulting estimates allowed us to use those chosen models as teammates for a second batch of models, which were used as teammates for a third batch of models, and so on, until all partition functions had been estimated. Because all but the initial estimates are computationally inexpensive, this method is overall much more efficient. (See the description of the results in section~\ref{sec: results} below.)

We chose the seed set as follows. For each of our 27 models $x$, we listed each other model $y$ for which $x$ was a predicted good teammate for $y$. For each $x$, and for each $y$ in the list for $x$, we then added to $x$'s list all models $z$ for which $y$ was also a predicted good teammate for $z$ (assuming $z$ was not already on the list). We did this iteratively until we reached a step where no additional models were added to the lists. The result of this is a list of ``descendants'' for each model. We chose the model with the most descendants as our first model in the seed set, breaking ties arbitrarily. We call this model $a$. In our data, we found that there was no one model that had every other model as a descendant. Therefore, for all the models $x$ which were not descendants of $a$, we listed the descendants of $x$ which were not descendants of $a$, and the one with the most such descendants was chosen as the second model for the seed set, which we refer to as model $b$. We chose a third model for the seed set using a similar method. The three models chosen for the seed set in this way were those labeled yY7aq, J3AmH, and O8QGE. Collectively, these had all other models as descendants. 

The seed-set models were assigned their partition function estimates computed using the previous method. We then iterated through the remaining models, iterating through direct descendants of the seed-set models first. For each model $x$, we listed the models that had already been assigned a partition function that were predicted by the random forest classifier to be good teammates for $x$. Using each of those models, we then used free energy perturbation to compute a partition function estimate for $x$. The median of these estimates was then assigned as the partition function for $x$, and $x$ was added to the list of models that had been assigned partition functions. Both these assigned partition function estimates and the partition function estimates computed using the non-immune repertoire teammate were then used to do maximum likelihood inference on sequences.

\subsection{Quality Testing Via Inference on Sequences}

For each model, we made an aggregate estimate of the partition function as follows. First we identified all other models which were classified by our random-forest classifier as giving good estimates as teammates for that model. Our aggregate estimate was then the median of all of these estimates. We refer to these estimates the \emph{median teammate} estimates. 

To test if the median teammate estimates were sufficiently accurate, we took 100 samples of 100 sequences from each model, each selected as an independent sub-sample of the 300,000-sequence sample used to compute expectation values in that code. For each of these 2,700 samples (27 models $\times$ 100 samples per model) we used maximum likelihood to guess which model it originated from. We did this using both the likelihood estimates from the bridge-sampled partition functions and the median teammate partition functions. We then compared the accuracies of both for identifying the correct source models, to see if there was any significant diminution in accuracy resulting from using the median teammate estimate partition functions rather than the bridge sampling partition functions.

\section{Results}\label{sec: results}

\begin{figure*}
    \centering
    \begin{subfigure}[b]{0.63\textwidth}
        \centering
        \caption{}
        \includegraphics[width=1.0\textwidth]{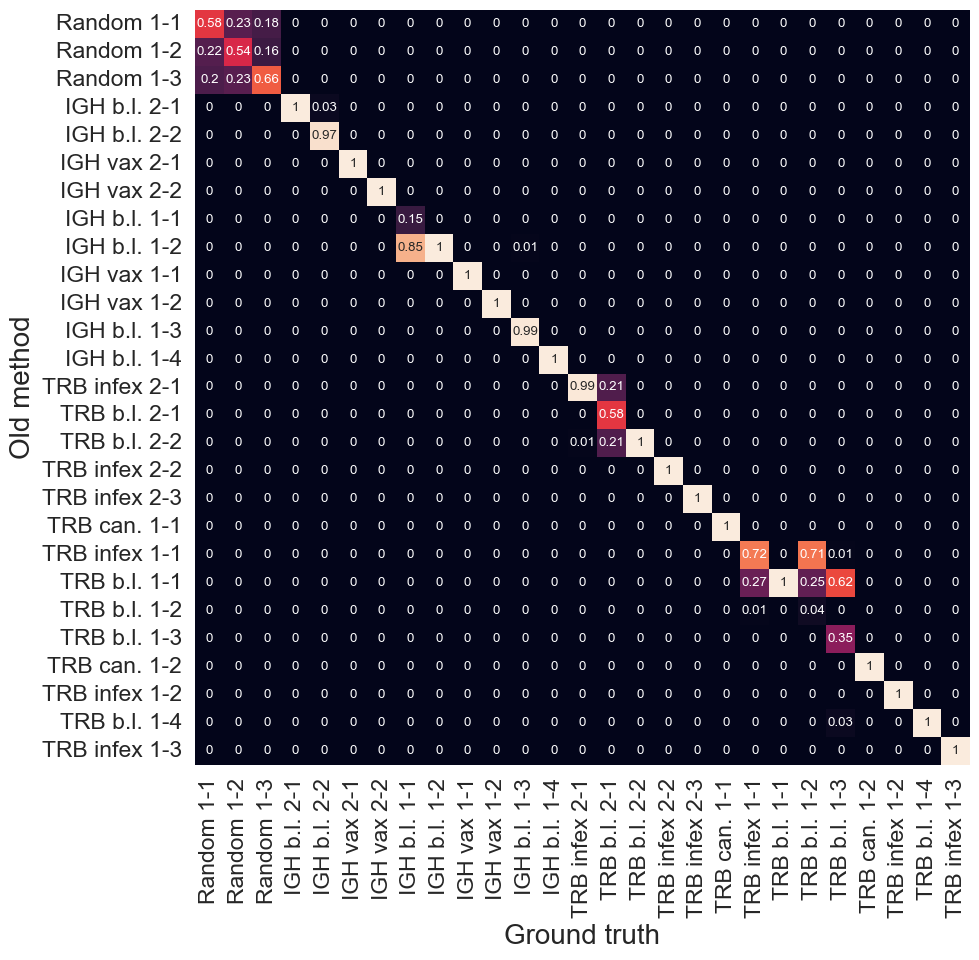}
    \end{subfigure}
    
    
    \begin{subfigure}[b]{0.63\textwidth}
        \centering
        \caption{}
        \includegraphics[width=1.0\textwidth]{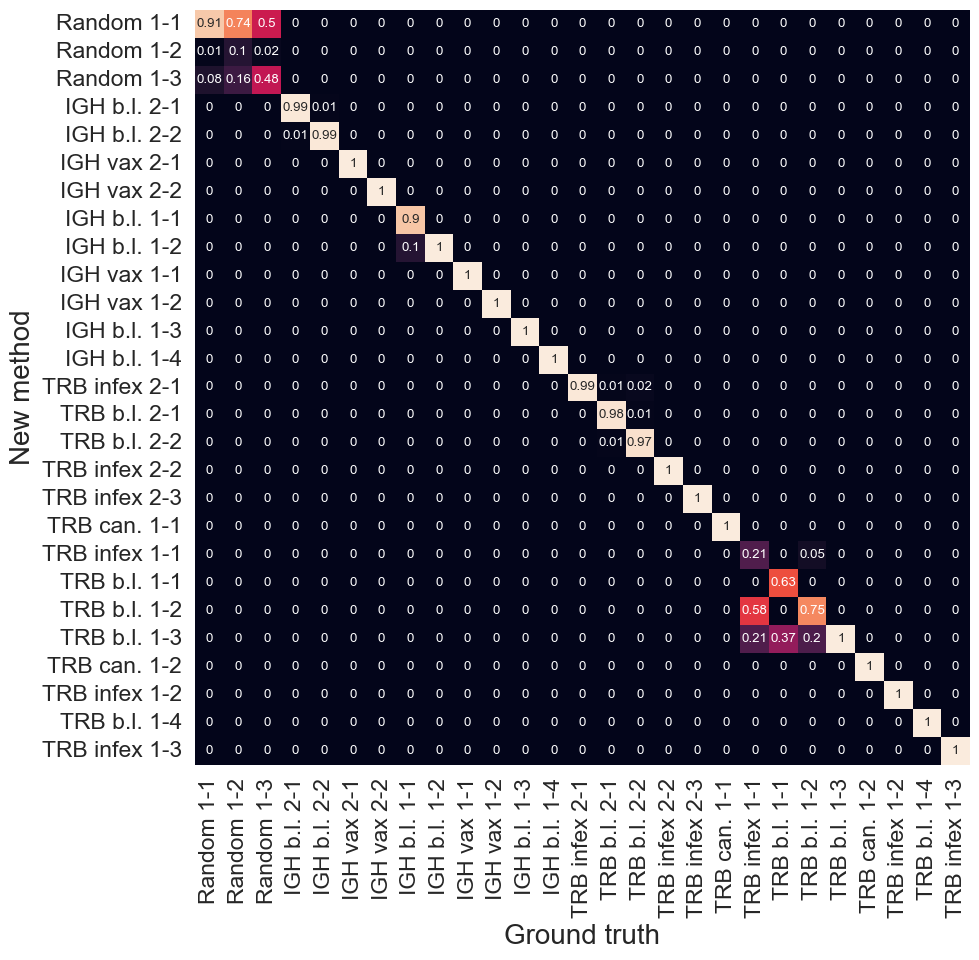}
    \end{subfigure}
    
    
    \centering
    \begin{subfigure}[b]{0.63\textwidth}
        \centering
        \includegraphics[width=1.0\textwidth]{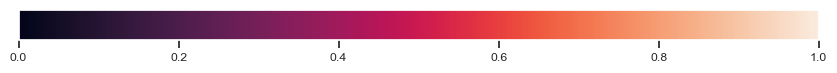}
    \end{subfigure}
    \caption{
        (a) Confusion matrices between models using method (\textit{i}).  (b) Confusion matrices between models using method (\textit{ii}). 
    }\label{fig: confusion_matrices}
\end{figure*}

\begin{figure*}
    \centering
    \begin{subfigure}[b]{0.48\textwidth}
        \centering
        \caption{}
        \includegraphics[width=1.0\textwidth]{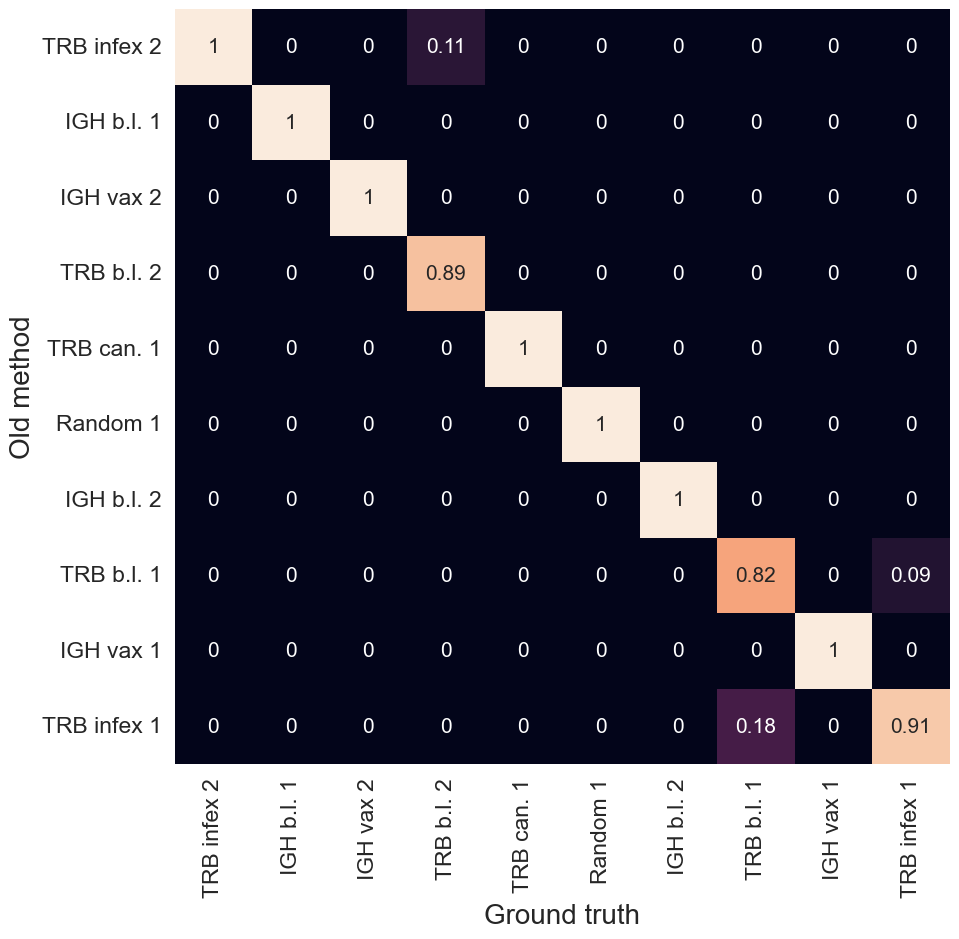}
    \end{subfigure}
    \hfill
    \begin{subfigure}[b]{0.48\textwidth}
        \centering
        \caption{}
        \includegraphics[width=1.0\textwidth]{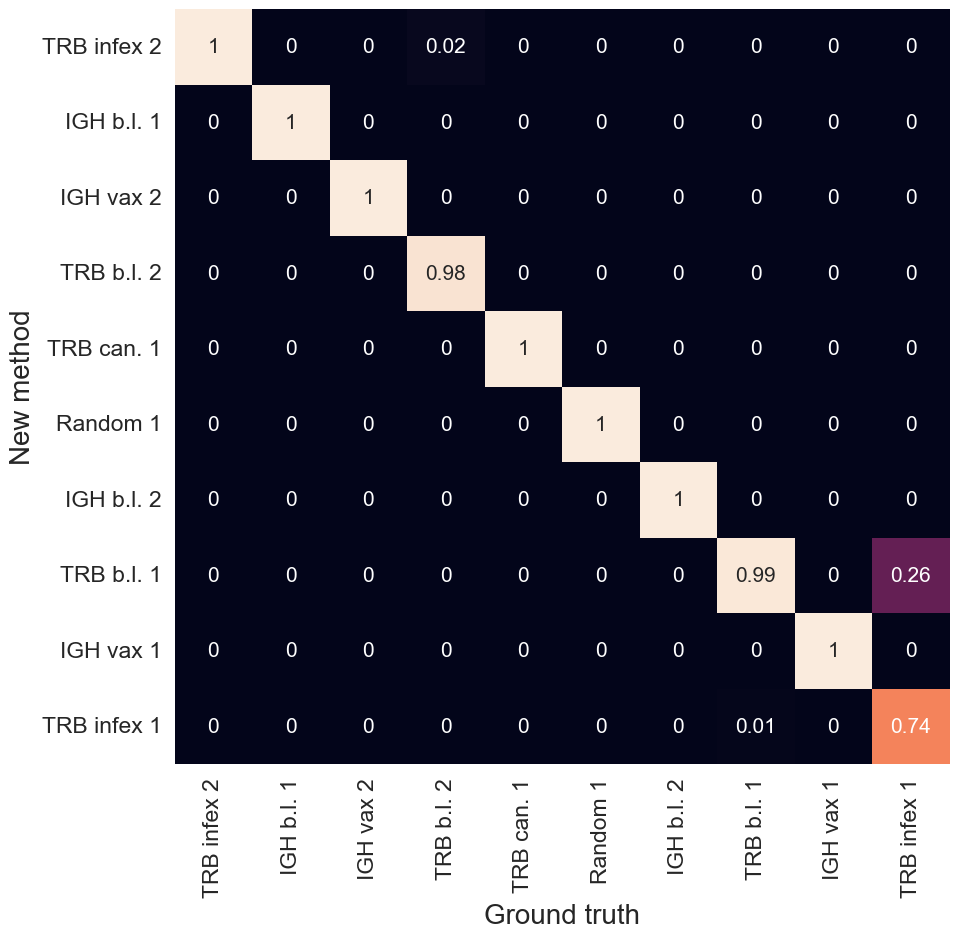}
    \end{subfigure}
    
    
    \centering
    \begin{subfigure}[b]{0.96\textwidth}
        \centering
        \includegraphics[width=1.0\textwidth]{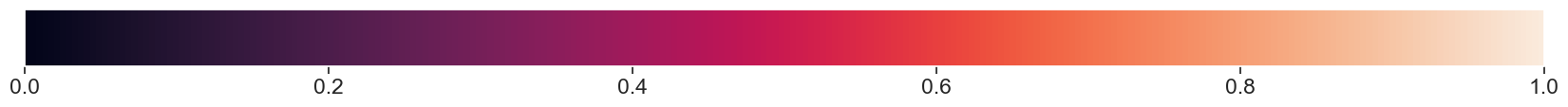}
    \end{subfigure}
    \caption{
        (a) Confusion matrices between model parameter set/cell type/disease state labels using method (\textit{i}). (b) Confusion matrices between model parameter set/cell type/disease state labels using method (\textit{ii}). b.l. = baseline, vax = vaccinated, infex = infected, can. = cancer, and 1,2 refers to the feature set.
    }\label{fig: confusion_matrices2}
\end{figure*}

We compared the previous method\cite{Arora519108} and the new method for both computational efficiency and performance of the resulting Bayesian classification.
Figures~\ref{fig: confusion_matrices} and~\ref{fig: confusion_matrices2} 
show the confusion matrices resulting from classifying 100 Monte-Carlo-generated sequences, both by the exact repertoire model they were sampled from (Fig.~\ref{fig: confusion_matrices}), as well as by sequence type (IGH vs. TRB) and disease or immunization state (Fig.~\ref{fig: confusion_matrices2}). 
Comparisons of Figs.~\ref{fig: confusion_matrices}a-~\ref{fig: confusion_matrices}b to Figs.~\ref{fig: confusion_matrices2}a-~\ref{fig: confusion_matrices2}b show that the new method gives comparable classification performance.

In addition, the new method had a much lower computational cost. Most of the cost came from the MC generation of sequence samples. Each estimate based on the non-immune repertoire teammates required about $10^{10}$ sequences. In contrast, the estimates using another immune repertoire model as a teammate only used $3\times10^5$ sequences per model, a savings of 99.997\%. This translates to about $10^7$ sequences in total across the 27 models---negligible compared to the previous method. As such, the computational cost essentially comes down to the number of more expensive estimates that need to be generated. In the previous method, each of the 27 estimates were of the expensive type. For the new method, we required only 3 of the expensive estimates, from which partition functions for the other 24 models were estimated much more efficiently. As a result, overall the new method leads to about an order of magnitude savings in computational cost.

\section{Discussion}\label{sec: discussion}

EBMs provide a way to summarize complex systems such as immune repertoires compactly and efficiently, based on aggregate features that are often human-interpretable. They also have the advantage of being generative models, which for immune repertoires means they can quickly and easily produce arbitrarily many \textit{de novo} sequences that are representative of a given repertoire. Partition function estimation is important in EBMs because it allows calculation of the absolute probability of a given state---for example determining which of several immunological states, such as infection or cancer (each described by one or more models), a set of sequences is diagnostically most consistent with~\cite{arnaout2021}. 
Together with methods for measuring immunological diversity, EBMs could become an important part of the diagnostic toolkit in next-generation immunology~\cite{kaplinsky2016, arora2022}, 
provided partition functions can be estimated efficiently. Here we have demonstrated a highly-efficient new method for estimating partition functions that performs as well as a previous method but much more efficiently, as assessed by correct classification of immune-repertoire sequences using models of real-world repertoires.

Although we have demonstrated substantial computational savings on this set of diverse IGH and TRB repertoires from a variety of states of health and disease, it should be noted that further work is necessary to precisely define how the computational cost of estimating partition functions will scale as the number of models requiring paritition-function estimation increases into the hundreds or thousands. The answer will likely depend in part on how closely related the models are, since we find closely related models tend to make good teammates. If the new method were to scale linearly, as the previous method does~\cite{Arora519108}, then the advantage would be merely a (substantial) multiplicative factor. Based on our results, the new method likely scales sub-linearly, significantly improving the utility of this method in situations where many repertoires are modeled, e.g. representing precisely defined or multifaceted disease states across large clinical cohorts. Curating a database of previously fitted models with partition functions computed would maximize the cost savings of this method, by creating a bank of potential teammates for use in normalizing new models.

In the new method presented here, after partition functions for the seed models are estimated, additional estimates are found using free energy perturbation, arguably the simplest of the MCMC-based methods. In the future, it may be interesting to implement this idea using other MCMC-based methods, to test more broadly how those estimates compare in terms of accuracy and computational cost. It would also be interesting to explore replacing the initial teammate used in the old method, for which probabilities depended only on sequence length, with a better choice of teammate. A model trained on the same repertoire as the target of interest, but with features restricted to contain only couplings between nearest-neighbor pairs of amino acids\footnote{Longer range interactions up to $k$th nearest neighbors can be included by grouping $k$ amino acids into a single variable with $20^k$ states, though the cost of the exact computation scales exponentially in $k$. }, would be formally the same as a 20-state 1D Potts model for each length, and as such the partition function for such a model could be found exactly using standard methods. This class of models could provide an improved teammate for the initial estimate, further reducing the cost to normalize an entire batch of models.

We conclude with a general note regarding obstacles to interdisciplinary adoption of EBMs.  
While the literature on MC methods for computing partition functions is extensive and goes back decades, it primarily traces its origins to statistics and  statistical physics.
Consequently, 
terminologies and concepts may not be readily accessible to researchers from diverse fields, including the biomedical sciences, whom they could otherwise benefit. Therefore, it may be useful to have a review that introduces these concepts specifically to biomedical researchers. Such a resource would facilitate the dissemination of knowledge, help avoid delays due to reinvention, and encourage the adoption of these powerful computational tools in various research domains. This is especially as the amount of data available in biology and related fields continues to increase, bringing ever-more-complex systems more fully into the realm of scientific study.


\hfill

\section{Acknowledgements}

The authors would like to acknowledge that the Research Computing group in the Division of Information Technology
at the University of South Carolina contributed to the results of this research by providing High Performance Computing resources and expertise.
This work was supported by the NIH (R01AI148747-01).




\bibliography{teamwork}

\begin{thebibliography}{39}%
\makeatletter
\providecommand \@ifxundefined [1]{%
 \@ifx{#1\undefined}
}%
\providecommand \@ifnum [1]{%
 \ifnum #1\expandafter \@firstoftwo
 \else \expandafter \@secondoftwo
 \fi
}%
\providecommand \@ifx [1]{%
 \ifx #1\expandafter \@firstoftwo
 \else \expandafter \@secondoftwo
 \fi
}%
\providecommand \natexlab [1]{#1}%
\providecommand \enquote  [1]{``#1''}%
\providecommand \bibnamefont  [1]{#1}%
\providecommand \bibfnamefont [1]{#1}%
\providecommand \citenamefont [1]{#1}%
\providecommand \href@noop [0]{\@secondoftwo}%
\providecommand \href [0]{\begingroup \@sanitize@url \@href}%
\providecommand \@href[1]{\@@startlink{#1}\@@href}%
\providecommand \@@href[1]{\endgroup#1\@@endlink}%
\providecommand \@sanitize@url [0]{\catcode `\\12\catcode `\$12\catcode
  `\&12\catcode `\#12\catcode `\^12\catcode `\_12\catcode `\%12\relax}%
\providecommand \@@startlink[1]{}%
\providecommand \@@endlink[0]{}%
\providecommand \url  [0]{\begingroup\@sanitize@url \@url }%
\providecommand \@url [1]{\endgroup\@href {#1}{\urlprefix }}%
\providecommand \urlprefix  [0]{URL }%
\providecommand \Eprint [0]{\href }%
\providecommand \doibase [0]{https://doi.org/}%
\providecommand \selectlanguage [0]{\@gobble}%
\providecommand \bibinfo  [0]{\@secondoftwo}%
\providecommand \bibfield  [0]{\@secondoftwo}%
\providecommand \translation [1]{[#1]}%
\providecommand \BibitemOpen [0]{}%
\providecommand \bibitemStop [0]{}%
\providecommand \bibitemNoStop [0]{.\EOS\space}%
\providecommand \EOS [0]{\spacefactor3000\relax}%
\providecommand \BibitemShut  [1]{\csname bibitem#1\endcsname}%
\let\auto@bib@innerbib\@empty
\bibitem [{\citenamefont {Boltzmann}(1868)}]{boltzmann1868}%
  \BibitemOpen
  \bibfield  {author} {\bibinfo {author} {\bibfnamefont {L.}~\bibnamefont
  {Boltzmann}},\ }\bibfield  {title} {\bibinfo {title} {Studien über das
  gleichgewicht der lebendigen kraft zwischen bewegten materiellen punkten},\
  }\href@noop {} {\bibfield  {journal} {\bibinfo  {journal} {Wiener Berichte}\
  }\textbf {\bibinfo {volume} {58}},\ \bibinfo {pages} {517} (\bibinfo {year}
  {1868})}\BibitemShut {NoStop}%
\bibitem [{\citenamefont {Gibbs}(1876)}]{gibbs1876}%
  \BibitemOpen
  \bibfield  {author} {\bibinfo {author} {\bibfnamefont {J.~W.}\ \bibnamefont
  {Gibbs}},\ }\bibfield  {title} {\bibinfo {title} {On the equilibrium of
  heterogeneous substances},\ }\href@noop {} {\bibfield  {journal} {\bibinfo
  {journal} {Transactions of the Connecticut Academy of Arts and Sciences}\
  }\textbf {\bibinfo {volume} {3}},\ \bibinfo {pages} {108–248} (\bibinfo
  {year} {October 1875 – May 1876})}\BibitemShut {NoStop}%
\bibitem [{\citenamefont {Gibbs}(1878)}]{gibbs1878}%
  \BibitemOpen
  \bibfield  {author} {\bibinfo {author} {\bibfnamefont {J.~W.}\ \bibnamefont
  {Gibbs}},\ }\bibfield  {title} {\bibinfo {title} {On the equilibrium of
  heterogeneous substances},\ }\href@noop {} {\bibfield  {journal} {\bibinfo
  {journal} {Transactions of the Connecticut Academy of Arts and Sciences}\
  }\textbf {\bibinfo {volume} {3}},\ \bibinfo {pages} {343–524} (\bibinfo
  {year} {May 1877 – July 1878})}\BibitemShut {NoStop}%
\bibitem [{\citenamefont {Jaynes}(1957{\natexlab{a}})}]{jaynes1957a}%
  \BibitemOpen
  \bibfield  {author} {\bibinfo {author} {\bibfnamefont {E.~T.}\ \bibnamefont
  {Jaynes}},\ }\bibfield  {title} {\bibinfo {title} {Information {Theory} and
  {Statistical} {Mechanics}},\ }\href {https://doi.org/10.1103/PhysRev.106.620}
  {\bibfield  {journal} {\bibinfo  {journal} {Physical Review}\ }\textbf
  {\bibinfo {volume} {106}},\ \bibinfo {pages} {620} (\bibinfo {year}
  {1957}{\natexlab{a}})}\BibitemShut {NoStop}%
\bibitem [{\citenamefont {Jaynes}(1957{\natexlab{b}})}]{jaynes1957b}%
  \BibitemOpen
  \bibfield  {author} {\bibinfo {author} {\bibfnamefont {E.~T.}\ \bibnamefont
  {Jaynes}},\ }\bibfield  {title} {\bibinfo {title} {Information {Theory} and
  {Statistical} {Mechanics}. {II}},\ }\href
  {https://doi.org/10.1103/PhysRev.108.171} {\bibfield  {journal} {\bibinfo
  {journal} {Physical Review}\ }\textbf {\bibinfo {volume} {108}},\ \bibinfo
  {pages} {171} (\bibinfo {year} {1957}{\natexlab{b}})}\BibitemShut {NoStop}%
\bibitem [{\citenamefont {Lafferty}\ and\ \citenamefont
  {Suhm}(1996)}]{lafferty1995}%
  \BibitemOpen
  \bibfield  {author} {\bibinfo {author} {\bibfnamefont {J.~D.}\ \bibnamefont
  {Lafferty}}\ and\ \bibinfo {author} {\bibfnamefont {B.}~\bibnamefont
  {Suhm}},\ }\bibfield  {title} {\bibinfo {title} {Cluster expansions and
  iterative scaling for maximum entropy language models},\ }in\ \href
  {http://arxiv.org/abs/cmp-lg/9509003} {\emph {\bibinfo {booktitle} {Maximum
  Entropy and Bayesian Methods}}},\ \bibinfo {series and number} {\bibinfo
  {number} {{arXiv}:cmp-lg/9509003}},\ \bibinfo {editor} {edited by\ \bibinfo
  {editor} {\bibfnamefont {K.~M.}\ \bibnamefont {Hanson}}\ and\ \bibinfo
  {editor} {\bibfnamefont {R.~N.}\ \bibnamefont {Silver}}}\ (\bibinfo
  {publisher} {Springer Netherlands},\ \bibinfo {address} {Dordrecht},\
  \bibinfo {year} {1996})\ pp.\ \bibinfo {pages} {195--202},\ \Eprint
  {https://arxiv.org/abs/cmp-lg/9509003} {cmp-lg/9509003} \BibitemShut
  {NoStop}%
\bibitem [{\citenamefont {Berger}\ \emph {et~al.}(1996)\citenamefont {Berger},
  \citenamefont {Della~Pietra},\ and\ \citenamefont
  {Della~Pietra}}]{berger1996}%
  \BibitemOpen
  \bibfield  {author} {\bibinfo {author} {\bibfnamefont {A.~L.}\ \bibnamefont
  {Berger}}, \bibinfo {author} {\bibfnamefont {S.~A.}\ \bibnamefont
  {Della~Pietra}},\ and\ \bibinfo {author} {\bibfnamefont {V.~J.}\ \bibnamefont
  {Della~Pietra}},\ }\bibfield  {title} {\bibinfo {title} {A maximum entropy
  approach to natural language processing},\ }\href
  {https://aclanthology.org/J96-1002} {\bibfield  {journal} {\bibinfo
  {journal} {Computational Linguistics}\ }\textbf {\bibinfo {volume} {22}},\
  \bibinfo {pages} {39} (\bibinfo {year} {1996})}\BibitemShut {NoStop}%
\bibitem [{\citenamefont {Molins}\ and\ \citenamefont
  {Vives}(2004)}]{molins2004}%
  \BibitemOpen
  \bibfield  {author} {\bibinfo {author} {\bibfnamefont {J.}~\bibnamefont
  {Molins}}\ and\ \bibinfo {author} {\bibfnamefont {E.}~\bibnamefont {Vives}},\
  }\href {http://arxiv.org/abs/cond-mat/0401378} {\bibinfo {title} {Long range
  ising model for credit risk modeling in homogeneous portfolios}} (\bibinfo
  {year} {2004}),\ \Eprint {https://arxiv.org/abs/cond-mat/0401378}
  {cond-mat/0401378} \BibitemShut {NoStop}%
\bibitem [{\citenamefont {Yeo}\ and\ \citenamefont {Burge}(2004)}]{yeo2004}%
  \BibitemOpen
  \bibfield  {author} {\bibinfo {author} {\bibfnamefont {G.}~\bibnamefont
  {Yeo}}\ and\ \bibinfo {author} {\bibfnamefont {C.~B.}\ \bibnamefont
  {Burge}},\ }\bibfield  {title} {\bibinfo {title} {Maximum entropy modeling of
  short sequence motifs with applications to {RNA} splicing signals},\ }\href
  {https://doi.org/10.1089/1066527041410418} {\bibfield  {journal} {\bibinfo
  {journal} {Journal of Computational Biology: A Journal of Computational
  Molecular Cell Biology}\ }\textbf {\bibinfo {volume} {11}},\ \bibinfo {pages}
  {377} (\bibinfo {year} {2004})}\BibitemShut {NoStop}%
\bibitem [{\citenamefont {Shimagaki}\ and\ \citenamefont
  {Weigt}(2019)}]{Shimagaki2019}%
  \BibitemOpen
  \bibfield  {author} {\bibinfo {author} {\bibfnamefont {K.}~\bibnamefont
  {Shimagaki}}\ and\ \bibinfo {author} {\bibfnamefont {M.}~\bibnamefont
  {Weigt}},\ }\bibfield  {title} {\bibinfo {title} {Selection of sequence
  motifs and generative hopfield-potts models for protein families},\ }\href
  {https://doi.org/10.1103/PhysRevE.100.032128} {\bibfield  {journal} {\bibinfo
   {journal} {Phys. Rev. E}\ }\textbf {\bibinfo {volume} {100}},\ \bibinfo
  {pages} {032128} (\bibinfo {year} {2019})},\ \Eprint
  {https://arxiv.org/abs/1905.11848} {arXiv:1905.11848 [q-bio.BM]} \BibitemShut
  {NoStop}%
\bibitem [{\citenamefont {Shipley}\ \emph {et~al.}(2006)\citenamefont
  {Shipley}, \citenamefont {Vile},\ and\ \citenamefont
  {Garnier}}]{shipley2006}%
  \BibitemOpen
  \bibfield  {author} {\bibinfo {author} {\bibfnamefont {B.}~\bibnamefont
  {Shipley}}, \bibinfo {author} {\bibfnamefont {D.}~\bibnamefont {Vile}},\ and\
  \bibinfo {author} {\bibfnamefont {E.}~\bibnamefont {Garnier}},\ }\bibfield
  {title} {\bibinfo {title} {From plant traits to plant communities: A
  statistical mechanistic approach to biodiversity},\ }\href
  {https://doi.org/10.1126/science.1131344} {\bibfield  {journal} {\bibinfo
  {journal} {Science}\ }\textbf {\bibinfo {volume} {314}},\ \bibinfo {pages}
  {812} (\bibinfo {year} {2006})}\BibitemShut {NoStop}%
\bibitem [{\citenamefont {Phillips}\ \emph {et~al.}(2006)\citenamefont
  {Phillips}, \citenamefont {Anderson},\ and\ \citenamefont
  {Schapire}}]{phillips2006}%
  \BibitemOpen
  \bibfield  {author} {\bibinfo {author} {\bibfnamefont {S.~J.}\ \bibnamefont
  {Phillips}}, \bibinfo {author} {\bibfnamefont {R.~P.}\ \bibnamefont
  {Anderson}},\ and\ \bibinfo {author} {\bibfnamefont {R.~E.}\ \bibnamefont
  {Schapire}},\ }\bibfield  {title} {\bibinfo {title} {Maximum entropy modeling
  of species geographic distributions},\ }\href
  {https://doi.org/10.1016/j.ecolmodel.2005.03.026} {\bibfield  {journal}
  {\bibinfo  {journal} {Ecological Modelling}\ }\textbf {\bibinfo {volume}
  {190}},\ \bibinfo {pages} {231} (\bibinfo {year} {2006})}\BibitemShut
  {NoStop}%
\bibitem [{\citenamefont {Williams}(2009)}]{williams2009}%
  \BibitemOpen
  \bibfield  {author} {\bibinfo {author} {\bibfnamefont {R.~J.}\ \bibnamefont
  {Williams}},\ }\href {http://arxiv.org/abs/0901.0976} {\bibinfo {title}
  {Simple {MaxEnt} models for food web degree distributions}} (\bibinfo {year}
  {2009}),\ \Eprint {https://arxiv.org/abs/0901.0976 [q-bio]} {0901.0976
  [q-bio]} \BibitemShut {NoStop}%
\bibitem [{\citenamefont {Cavagna}\ \emph {et~al.}(2014)\citenamefont
  {Cavagna}, \citenamefont {Giardina}, \citenamefont {Ginelli}, \citenamefont
  {Mora}, \citenamefont {Piovani}, \citenamefont {Tavarone},\ and\
  \citenamefont {Walczak}}]{cavagna2014}%
  \BibitemOpen
  \bibfield  {author} {\bibinfo {author} {\bibfnamefont {A.}~\bibnamefont
  {Cavagna}}, \bibinfo {author} {\bibfnamefont {I.}~\bibnamefont {Giardina}},
  \bibinfo {author} {\bibfnamefont {F.}~\bibnamefont {Ginelli}}, \bibinfo
  {author} {\bibfnamefont {T.}~\bibnamefont {Mora}}, \bibinfo {author}
  {\bibfnamefont {D.}~\bibnamefont {Piovani}}, \bibinfo {author} {\bibfnamefont
  {R.}~\bibnamefont {Tavarone}},\ and\ \bibinfo {author} {\bibfnamefont
  {A.~M.}\ \bibnamefont {Walczak}},\ }\bibfield  {title} {\bibinfo {title}
  {Dynamical maximum entropy approach to flocking},\ }\href
  {https://doi.org/10.1103/PhysRevE.89.042707} {\bibfield  {journal} {\bibinfo
  {journal} {Physical Review E}\ }\textbf {\bibinfo {volume} {89}},\ \bibinfo
  {pages} {042707} (\bibinfo {year} {2014})},\ \Eprint
  {https://arxiv.org/abs/1310.3810 [cond-mat, physics:physics, q-bio]}
  {1310.3810 [cond-mat, physics:physics, q-bio]} \BibitemShut {NoStop}%
\bibitem [{\citenamefont {Lee}\ \emph {et~al.}(2015)\citenamefont {Lee},
  \citenamefont {Broedersz},\ and\ \citenamefont {Bialek}}]{lee2015}%
  \BibitemOpen
  \bibfield  {author} {\bibinfo {author} {\bibfnamefont {E.~D.}\ \bibnamefont
  {Lee}}, \bibinfo {author} {\bibfnamefont {C.~P.}\ \bibnamefont {Broedersz}},\
  and\ \bibinfo {author} {\bibfnamefont {W.}~\bibnamefont {Bialek}},\
  }\bibfield  {title} {\bibinfo {title} {Statistical mechanics of the {US}
  supreme court},\ }\href {https://doi.org/10.1007/s10955-015-1253-6}
  {\bibfield  {journal} {\bibinfo  {journal} {Journal of Statistical Physics}\
  }\textbf {\bibinfo {volume} {160}},\ \bibinfo {pages} {275} (\bibinfo {year}
  {2015})},\ \Eprint {https://arxiv.org/abs/1306.5004} {1306.5004} \BibitemShut
  {NoStop}%
\bibitem [{\citenamefont {Ferrari}\ \emph {et~al.}(2017)\citenamefont
  {Ferrari}, \citenamefont {Obuchi},\ and\ \citenamefont {Mora}}]{ferrari2017}%
  \BibitemOpen
  \bibfield  {author} {\bibinfo {author} {\bibfnamefont {U.}~\bibnamefont
  {Ferrari}}, \bibinfo {author} {\bibfnamefont {T.}~\bibnamefont {Obuchi}},\
  and\ \bibinfo {author} {\bibfnamefont {T.}~\bibnamefont {Mora}},\ }\bibfield
  {title} {\bibinfo {title} {Random versus maximum entropy models of neural
  population activity},\ }\href {https://doi.org/10.1103/PhysRevE.95.042321}
  {\bibfield  {journal} {\bibinfo  {journal} {Physical Review E}\ }\textbf
  {\bibinfo {volume} {95}},\ \bibinfo {pages} {042321} (\bibinfo {year}
  {2017})},\ \Eprint {https://arxiv.org/abs/1612.02807 [cond-mat, q-bio]}
  {1612.02807 [cond-mat, q-bio]} \BibitemShut {NoStop}%
\bibitem [{\citenamefont {Nghiem}\ \emph {et~al.}(2018)\citenamefont {Nghiem},
  \citenamefont {Telenczuk}, \citenamefont {Marre}, \citenamefont {Destexhe},\
  and\ \citenamefont {Ferrari}}]{Nghiem2018}%
  \BibitemOpen
  \bibfield  {author} {\bibinfo {author} {\bibfnamefont {T.-A.}\ \bibnamefont
  {Nghiem}}, \bibinfo {author} {\bibfnamefont {B.}~\bibnamefont {Telenczuk}},
  \bibinfo {author} {\bibfnamefont {O.}~\bibnamefont {Marre}}, \bibinfo
  {author} {\bibfnamefont {A.}~\bibnamefont {Destexhe}},\ and\ \bibinfo
  {author} {\bibfnamefont {U.}~\bibnamefont {Ferrari}},\ }\bibfield  {title}
  {\bibinfo {title} {Maximum-entropy models reveal the excitatory and
  inhibitory correlation structures in cortical neuronal activity},\ }\href
  {https://doi.org/10.1103/PhysRevE.98.012402} {\bibfield  {journal} {\bibinfo
  {journal} {Phys. Rev. E}\ }\textbf {\bibinfo {volume} {98}},\ \bibinfo
  {pages} {012402} (\bibinfo {year} {2018})},\ \Eprint
  {https://arxiv.org/abs/1801.01853} {arXiv:1801.01853 [q-bio.NC]} \BibitemShut
  {NoStop}%
\bibitem [{\citenamefont {Ansari}\ \emph {et~al.}(2022)\citenamefont {Ansari},
  \citenamefont {Soriano-Pa\~nos}, \citenamefont {Ghoshal},\ and\ \citenamefont
  {White}}]{Ansari2022}%
  \BibitemOpen
  \bibfield  {author} {\bibinfo {author} {\bibfnamefont {M.}~\bibnamefont
  {Ansari}}, \bibinfo {author} {\bibfnamefont {D.}~\bibnamefont
  {Soriano-Pa\~nos}}, \bibinfo {author} {\bibfnamefont {G.}~\bibnamefont
  {Ghoshal}},\ and\ \bibinfo {author} {\bibfnamefont {A.~D.}\ \bibnamefont
  {White}},\ }\bibfield  {title} {\bibinfo {title} {Inferring spatial source of
  disease outbreaks using maximum entropy},\ }\href
  {https://doi.org/10.1103/PhysRevE.106.014306} {\bibfield  {journal} {\bibinfo
   {journal} {Phys. Rev. E}\ }\textbf {\bibinfo {volume} {106}},\ \bibinfo
  {pages} {014306} (\bibinfo {year} {2022})},\ \Eprint
  {https://arxiv.org/abs/2110.03846} {arXiv:2110.03846 [physics.soc-ph]}
  \BibitemShut {NoStop}%
\bibitem [{\citenamefont {Cohen}(2023)}]{cohen2023}%
  \BibitemOpen
  \bibfield  {author} {\bibinfo {author} {\bibfnamefont {S.~D.}\ \bibnamefont
  {Cohen}},\ }\bibfield  {title} {\bibinfo {title} {Estimating the climate
  niche of sclerotinia sclerotiorum using maximum entropy modeling},\ }\href
  {https://doi.org/10.3390/jof9090892} {\bibfield  {journal} {\bibinfo
  {journal} {Journal of Fungi (Basel, Switzerland)}\ }\textbf {\bibinfo
  {volume} {9}},\ \bibinfo {pages} {892} (\bibinfo {year} {2023})}\BibitemShut
  {NoStop}%
\bibitem [{\citenamefont {Mora}\ \emph {et~al.}(2010)\citenamefont {Mora},
  \citenamefont {Walczak}, \citenamefont {Bialek},\ and\ \citenamefont
  {Callan}}]{mora_maximum_2010}%
  \BibitemOpen
  \bibfield  {author} {\bibinfo {author} {\bibfnamefont {T.}~\bibnamefont
  {Mora}}, \bibinfo {author} {\bibfnamefont {A.~M.}\ \bibnamefont {Walczak}},
  \bibinfo {author} {\bibfnamefont {W.}~\bibnamefont {Bialek}},\ and\ \bibinfo
  {author} {\bibfnamefont {C.~G.}\ \bibnamefont {Callan}},\ }\bibfield  {title}
  {\bibinfo {title} {Maximum entropy models for antibody diversity},\ }\href
  {https://doi.org/10.1073/pnas.1001705107} {\bibfield  {journal} {\bibinfo
  {journal} {Proceedings of the National Academy of Sciences of the United
  States of America}\ }\textbf {\bibinfo {volume} {107}},\ \bibinfo {pages}
  {5405} (\bibinfo {year} {2010})}\BibitemShut {NoStop}%
\bibitem [{\citenamefont {Arora}\ \emph {et~al.}(2019)\citenamefont {Arora},
  \citenamefont {Kaplinsky}, \citenamefont {Li},\ and\ \citenamefont
  {Arnaout}}]{Arora519108}%
  \BibitemOpen
  \bibfield  {author} {\bibinfo {author} {\bibfnamefont {R.}~\bibnamefont
  {Arora}}, \bibinfo {author} {\bibfnamefont {J.}~\bibnamefont {Kaplinsky}},
  \bibinfo {author} {\bibfnamefont {A.}~\bibnamefont {Li}},\ and\ \bibinfo
  {author} {\bibfnamefont {R.}~\bibnamefont {Arnaout}},\ }\bibfield  {title}
  {\bibinfo {title} {Repertoire-based diagnostics using statistical
  biophysics},\ }\bibfield  {journal} {\bibinfo  {journal} {bioRxiv}\ }\href
  {https://doi.org/10.1101/519108} {10.1101/519108} (\bibinfo {year} {2019}),\
  \Eprint
  {https://arxiv.org/abs/https://www.biorxiv.org/content/early/2019/01/13/519108}
  {https://www.biorxiv.org/content/early/2019/01/13/519108} \BibitemShut
  {NoStop}%
\bibitem [{\citenamefont {De~Martino}\ and\ \citenamefont
  {De~Martino}(2018)}]{de_martino2018}%
  \BibitemOpen
  \bibfield  {author} {\bibinfo {author} {\bibfnamefont {A.}~\bibnamefont
  {De~Martino}}\ and\ \bibinfo {author} {\bibfnamefont {D.}~\bibnamefont
  {De~Martino}},\ }\bibfield  {title} {\bibinfo {title} {An introduction to the
  maximum entropy approach and its application to inference problems in
  biology},\ }\href {https://doi.org/10.1016/j.heliyon.2018.e00596} {\bibfield
  {journal} {\bibinfo  {journal} {Heliyon}\ }\textbf {\bibinfo {volume} {4}},\
  \bibinfo {pages} {e00596} (\bibinfo {year} {2018})}\BibitemShut {NoStop}%
\bibitem [{\citenamefont {Agrawal}\ \emph {et~al.}(2021)\citenamefont
  {Agrawal}, \citenamefont {Pote},\ and\ \citenamefont {Meel}}]{agrawal2021}%
  \BibitemOpen
  \bibfield  {author} {\bibinfo {author} {\bibfnamefont {D.}~\bibnamefont
  {Agrawal}}, \bibinfo {author} {\bibfnamefont {Y.}~\bibnamefont {Pote}},\ and\
  \bibinfo {author} {\bibfnamefont {K.~S.}\ \bibnamefont {Meel}},\ }\href
  {http://arxiv.org/abs/2105.11132} {\bibinfo {title} {Partition {Function}
  {Estimation}: {A} {Quantitative} {Study}}} (\bibinfo {year} {2021}),\ \Eprint
  {https://arxiv.org/abs/2105.11132} {arXiv:2105.11132 [cs.AI]} \BibitemShut
  {NoStop}%
\bibitem [{\citenamefont {Roth}(1996)}]{roth1996}%
  \BibitemOpen
  \bibfield  {author} {\bibinfo {author} {\bibfnamefont {D.}~\bibnamefont
  {Roth}},\ }\bibfield  {title} {\bibinfo {title} {On the hardness of
  approximate reasoning},\ }\href
  {https://doi.org/10.1016/0004-3702(94)00092-1} {\bibfield  {journal}
  {\bibinfo  {journal} {Artificial Intelligence}\ }\textbf {\bibinfo {volume}
  {82}},\ \bibinfo {pages} {273} (\bibinfo {year} {1996})}\BibitemShut
  {NoStop}%
\bibitem [{\citenamefont {Bennett}(1976)}]{bennett1976}%
  \BibitemOpen
  \bibfield  {author} {\bibinfo {author} {\bibfnamefont {C.~H.}\ \bibnamefont
  {Bennett}},\ }\bibfield  {title} {\bibinfo {title} {Efficient estimation of
  free energy differences from monte carlo data},\ }\href
  {https://doi.org/https://doi.org/10.1016/0021-9991(76)90078-4} {\bibfield
  {journal} {\bibinfo  {journal} {Journal of Computational Physics}\ }\textbf
  {\bibinfo {volume} {22}},\ \bibinfo {pages} {245} (\bibinfo {year}
  {1976})}\BibitemShut {NoStop}%
\bibitem [{\citenamefont {MENG X.~L.}(1996)}]{Meng1996}%
  \BibitemOpen
  \bibfield  {author} {\bibinfo {author} {\bibfnamefont {W.~W.~H.}\
  \bibnamefont {MENG X.~L.}},\ }\bibfield  {title} {\bibinfo {title}
  {Simulating ratios of normalizing constants via a simple identity: a
  theoretical exploration},\ }\href
  {https://cir.nii.ac.jp/crid/1571417124797419520} {\bibfield  {journal}
  {\bibinfo  {journal} {Statistica Sinica}\ }\textbf {\bibinfo {volume} {6}},\
  \bibinfo {pages} {831} (\bibinfo {year} {1996})}\BibitemShut {NoStop}%
\bibitem [{\citenamefont {Gronau}\ \emph {et~al.}(2017)\citenamefont {Gronau},
  \citenamefont {Sarafoglou}, \citenamefont {Matzke}, \citenamefont {Ly},
  \citenamefont {Boehm}, \citenamefont {Marsman}, \citenamefont {Leslie},
  \citenamefont {Forster}, \citenamefont {Wagenmakers},\ and\ \citenamefont
  {Steingroever}}]{Gronau2017}%
  \BibitemOpen
  \bibfield  {author} {\bibinfo {author} {\bibfnamefont {Q.~F.}\ \bibnamefont
  {Gronau}}, \bibinfo {author} {\bibfnamefont {A.}~\bibnamefont {Sarafoglou}},
  \bibinfo {author} {\bibfnamefont {D.}~\bibnamefont {Matzke}}, \bibinfo
  {author} {\bibfnamefont {A.}~\bibnamefont {Ly}}, \bibinfo {author}
  {\bibfnamefont {U.}~\bibnamefont {Boehm}}, \bibinfo {author} {\bibfnamefont
  {M.}~\bibnamefont {Marsman}}, \bibinfo {author} {\bibfnamefont {D.~S.}\
  \bibnamefont {Leslie}}, \bibinfo {author} {\bibfnamefont {J.~J.}\
  \bibnamefont {Forster}}, \bibinfo {author} {\bibfnamefont {E.-J.}\
  \bibnamefont {Wagenmakers}},\ and\ \bibinfo {author} {\bibfnamefont
  {H.}~\bibnamefont {Steingroever}},\ }\href {http://arxiv.org/abs/1703.05984}
  {\bibinfo {title} {A {Tutorial} on {Bridge} {Sampling}}} (\bibinfo {year}
  {2017}),\ \Eprint {https://arxiv.org/abs/1703.05984} {arXiv:1703.05984
  [stat.CO]} \BibitemShut {NoStop}%
\bibitem [{\citenamefont {Russ}\ \emph {et~al.}(2005)\citenamefont {Russ},
  \citenamefont {Lowery}, \citenamefont {Mishra}, \citenamefont {Yaffe},\ and\
  \citenamefont {Ranganathan}}]{russ_natural-like_2005}%
  \BibitemOpen
  \bibfield  {author} {\bibinfo {author} {\bibfnamefont {W.~P.}\ \bibnamefont
  {Russ}}, \bibinfo {author} {\bibfnamefont {D.~M.}\ \bibnamefont {Lowery}},
  \bibinfo {author} {\bibfnamefont {P.}~\bibnamefont {Mishra}}, \bibinfo
  {author} {\bibfnamefont {M.~B.}\ \bibnamefont {Yaffe}},\ and\ \bibinfo
  {author} {\bibfnamefont {R.}~\bibnamefont {Ranganathan}},\ }\bibfield
  {title} {\bibinfo {title} {Natural-like function in artificial {WW}
  domains},\ }\href {https://doi.org/10.1038/nature03990} {\bibfield  {journal}
  {\bibinfo  {journal} {Nature}\ }\textbf {\bibinfo {volume} {437}},\ \bibinfo
  {pages} {579} (\bibinfo {year} {2005})}\BibitemShut {NoStop}%
\bibitem [{\citenamefont {Neal}(1993)}]{Neal1993}%
  \BibitemOpen
  \bibfield  {author} {\bibinfo {author} {\bibfnamefont {R.~M.}\ \bibnamefont
  {Neal}},\ }\href@noop {} {\emph {\bibinfo {title} {Probabilistic Inference
  Using Markov Chain Monte Carlo Methods}}},\ \bibinfo {type} {Tech. Rep.}\
  (\bibinfo  {institution} {Dept. of Computer Science, University of
  Toronto,},\ \bibinfo {year} {1993})\BibitemShut {NoStop}%
\bibitem [{\citenamefont {Zwanzig}(1954)}]{zwanzig1954}%
  \BibitemOpen
  \bibfield  {author} {\bibinfo {author} {\bibfnamefont {R.~W.}\ \bibnamefont
  {Zwanzig}},\ }\bibfield  {title} {\bibinfo {title} {High-{Temperature}
  {Equation} of {State} by a {Perturbation} {Method}. {I}. {Nonpolar}
  {Gases}},\ }\href {https://doi.org/10.1063/1.1740409} {\bibfield  {journal}
  {\bibinfo  {journal} {The Journal of Chemical Physics}\ }\textbf {\bibinfo
  {volume} {22}},\ \bibinfo {pages} {1420} (\bibinfo {year}
  {1954})}\BibitemShut {NoStop}%
\bibitem [{\citenamefont {Geyer}\ and\ \citenamefont
  {Thompson}(1992)}]{geyer1992}%
  \BibitemOpen
  \bibfield  {author} {\bibinfo {author} {\bibfnamefont {C.~J.}\ \bibnamefont
  {Geyer}}\ and\ \bibinfo {author} {\bibfnamefont {E.~A.}\ \bibnamefont
  {Thompson}},\ }\bibfield  {title} {\bibinfo {title} {Constrained {Monte}
  {Carlo} {Maximum} {Likelihood} for {Dependent} {Data}},\ }\href
  {https://doi.org/10.1111/j.2517-6161.1992.tb01443.x} {\bibfield  {journal}
  {\bibinfo  {journal} {Journal of the Royal Statistical Society: Series B
  (Methodological)}\ }\textbf {\bibinfo {volume} {54}},\ \bibinfo {pages} {657}
  (\bibinfo {year} {1992})}\BibitemShut {NoStop}%
\bibitem [{\citenamefont {Neal}(2005)}]{neal2005}%
  \BibitemOpen
  \bibfield  {author} {\bibinfo {author} {\bibfnamefont {R.~M.}\ \bibnamefont
  {Neal}},\ }\href {http://arxiv.org/abs/math/0511216} {\bibinfo {title}
  {Estimating {Ratios} of {Normalizing} {Constants} {Using} {Linked}
  {Importance} {Sampling}}} (\bibinfo {year} {2005}),\ \Eprint
  {https://arxiv.org/abs/math/0511216} {arXiv:math/0511216 [math.ST]}
  \BibitemShut {NoStop}%
\bibitem [{\citenamefont {Britanova}\ \emph {et~al.}(2014)\citenamefont
  {Britanova}, \citenamefont {Putintseva}, \citenamefont {Shugay},
  \citenamefont {Merzlyak}, \citenamefont {Turchaninova}, \citenamefont
  {Staroverov}, \citenamefont {Bolotin}, \citenamefont {Lukyanov},
  \citenamefont {Bogdanova}, \citenamefont {Mamedov}, \citenamefont {Lebedev},\
  and\ \citenamefont {Chudakov}}]{britanova2014}%
  \BibitemOpen
  \bibfield  {author} {\bibinfo {author} {\bibfnamefont {O.~V.}\ \bibnamefont
  {Britanova}}, \bibinfo {author} {\bibfnamefont {E.~V.}\ \bibnamefont
  {Putintseva}}, \bibinfo {author} {\bibfnamefont {M.}~\bibnamefont {Shugay}},
  \bibinfo {author} {\bibfnamefont {E.~M.}\ \bibnamefont {Merzlyak}}, \bibinfo
  {author} {\bibfnamefont {M.~A.}\ \bibnamefont {Turchaninova}}, \bibinfo
  {author} {\bibfnamefont {D.~B.}\ \bibnamefont {Staroverov}}, \bibinfo
  {author} {\bibfnamefont {D.~A.}\ \bibnamefont {Bolotin}}, \bibinfo {author}
  {\bibfnamefont {S.}~\bibnamefont {Lukyanov}}, \bibinfo {author}
  {\bibfnamefont {E.~A.}\ \bibnamefont {Bogdanova}}, \bibinfo {author}
  {\bibfnamefont {I.~Z.}\ \bibnamefont {Mamedov}}, \bibinfo {author}
  {\bibfnamefont {Y.~B.}\ \bibnamefont {Lebedev}},\ and\ \bibinfo {author}
  {\bibfnamefont {D.~M.}\ \bibnamefont {Chudakov}},\ }\bibfield  {title}
  {\bibinfo {title} {Age-related decrease in {TCR} repertoire diversity
  measured with deep and normalized sequence profiling},\ }\href
  {https://doi.org/10.4049/jimmunol.1302064} {\bibfield  {journal} {\bibinfo
  {journal} {Journal of Immunology (Baltimore, Md.: 1950)}\ }\textbf {\bibinfo
  {volume} {192}},\ \bibinfo {pages} {2689} (\bibinfo {year}
  {2014})}\BibitemShut {NoStop}%
\bibitem [{\citenamefont {Beausang}\ \emph {et~al.}(2017)\citenamefont
  {Beausang}, \citenamefont {Wheeler}, \citenamefont {Chan}, \citenamefont
  {Hanft}, \citenamefont {Dirbas}, \citenamefont {Jeffrey},\ and\ \citenamefont
  {Quake}}]{beausang2017}%
  \BibitemOpen
  \bibfield  {author} {\bibinfo {author} {\bibfnamefont {J.~F.}\ \bibnamefont
  {Beausang}}, \bibinfo {author} {\bibfnamefont {A.~J.}\ \bibnamefont
  {Wheeler}}, \bibinfo {author} {\bibfnamefont {N.~H.}\ \bibnamefont {Chan}},
  \bibinfo {author} {\bibfnamefont {V.~R.}\ \bibnamefont {Hanft}}, \bibinfo
  {author} {\bibfnamefont {F.~M.}\ \bibnamefont {Dirbas}}, \bibinfo {author}
  {\bibfnamefont {S.~S.}\ \bibnamefont {Jeffrey}},\ and\ \bibinfo {author}
  {\bibfnamefont {S.~R.}\ \bibnamefont {Quake}},\ }\bibfield  {title} {\bibinfo
  {title} {T cell receptor sequencing of early-stage breast cancer tumors
  identifies altered clonal structure of the t cell repertoire},\ }\bibfield
  {journal} {\bibinfo  {journal} {Proceedings of the National Academy of
  Sciences}\ }\textbf {\bibinfo {volume} {114}},\ \href
  {https://doi.org/10.1073/pnas.1713863114} {10.1073/pnas.1713863114} (\bibinfo
  {year} {2017})\BibitemShut {NoStop}%
\bibitem [{\citenamefont {Arora}\ and\ \citenamefont
  {Arnaout}(2022)}]{arora2022}%
  \BibitemOpen
  \bibfield  {author} {\bibinfo {author} {\bibfnamefont {R.}~\bibnamefont
  {Arora}}\ and\ \bibinfo {author} {\bibfnamefont {R.}~\bibnamefont
  {Arnaout}},\ }\bibfield  {title} {\bibinfo {title} {Repertoire-scale measures
  of antigen binding},\ }\href {https://doi.org/10.1073/pnas.2203505119}
  {\bibfield  {journal} {\bibinfo  {journal} {Proceedings of the National
  Academy of Sciences of the United States of America}\ }\textbf {\bibinfo
  {volume} {119}},\ \bibinfo {pages} {e2203505119} (\bibinfo {year}
  {2022})}\BibitemShut {NoStop}%
\bibitem [{\citenamefont {Vollmers}\ \emph {et~al.}(2013)\citenamefont
  {Vollmers}, \citenamefont {Sit}, \citenamefont {Weinstein}, \citenamefont
  {Dekker},\ and\ \citenamefont {Quake}}]{vollmers2013}%
  \BibitemOpen
  \bibfield  {author} {\bibinfo {author} {\bibfnamefont {C.}~\bibnamefont
  {Vollmers}}, \bibinfo {author} {\bibfnamefont {R.~V.}\ \bibnamefont {Sit}},
  \bibinfo {author} {\bibfnamefont {J.~A.}\ \bibnamefont {Weinstein}}, \bibinfo
  {author} {\bibfnamefont {C.~L.}\ \bibnamefont {Dekker}},\ and\ \bibinfo
  {author} {\bibfnamefont {S.~R.}\ \bibnamefont {Quake}},\ }\bibfield  {title}
  {\bibinfo {title} {Genetic measurement of memory b-cell recall using antibody
  repertoire sequencing},\ }\href {https://doi.org/10.1073/pnas.1312146110}
  {\bibfield  {journal} {\bibinfo  {journal} {Proceedings of the National
  Academy of Sciences of the United States of America}\ }\textbf {\bibinfo
  {volume} {110}},\ \bibinfo {pages} {13463} (\bibinfo {year}
  {2013})}\BibitemShut {NoStop}%
\bibitem [{\citenamefont {Bashford-Rogers}\ \emph {et~al.}(2013)\citenamefont
  {Bashford-Rogers}, \citenamefont {Palser}, \citenamefont {Huntly},
  \citenamefont {Rance}, \citenamefont {Vassiliou}, \citenamefont {Follows},\
  and\ \citenamefont {Kellam}}]{bashford-rogers2013}%
  \BibitemOpen
  \bibfield  {author} {\bibinfo {author} {\bibfnamefont {R.~J.}\ \bibnamefont
  {Bashford-Rogers}}, \bibinfo {author} {\bibfnamefont {A.~L.}\ \bibnamefont
  {Palser}}, \bibinfo {author} {\bibfnamefont {B.~J.}\ \bibnamefont {Huntly}},
  \bibinfo {author} {\bibfnamefont {R.}~\bibnamefont {Rance}}, \bibinfo
  {author} {\bibfnamefont {G.~S.}\ \bibnamefont {Vassiliou}}, \bibinfo {author}
  {\bibfnamefont {G.~A.}\ \bibnamefont {Follows}},\ and\ \bibinfo {author}
  {\bibfnamefont {P.}~\bibnamefont {Kellam}},\ }\bibfield  {title} {\bibinfo
  {title} {Network properties derived from deep sequencing of human b-cell
  receptor repertoires delineate b-cell populations},\ }\href
  {https://doi.org/10.1101/gr.154815.113} {\bibfield  {journal} {\bibinfo
  {journal} {Genome Research}\ }\textbf {\bibinfo {volume} {23}},\ \bibinfo
  {pages} {1874} (\bibinfo {year} {2013})}\BibitemShut {NoStop}%
\bibitem [{\citenamefont {Arnaout}\ \emph {et~al.}(2021)\citenamefont
  {Arnaout}, \citenamefont {Prak}, \citenamefont {Schwab}, \citenamefont
  {Rubelt},\ and\ \citenamefont {{the Adaptive Immune Receptor Repertoire
  Community}}}]{arnaout2021}%
  \BibitemOpen
  \bibfield  {author} {\bibinfo {author} {\bibfnamefont {R.~A.}\ \bibnamefont
  {Arnaout}}, \bibinfo {author} {\bibfnamefont {E.~T.~L.}\ \bibnamefont
  {Prak}}, \bibinfo {author} {\bibfnamefont {N.}~\bibnamefont {Schwab}},
  \bibinfo {author} {\bibfnamefont {F.}~\bibnamefont {Rubelt}},\ and\ \bibinfo
  {author} {\bibnamefont {{the Adaptive Immune Receptor Repertoire
  Community}}},\ }\bibfield  {title} {\bibinfo {title} {The future of blood
  testing is the immunome},\ }\href {https://doi.org/10.3389/fimmu.2021.626793}
  {\bibfield  {journal} {\bibinfo  {journal} {Frontiers in Immunology}\
  }\textbf {\bibinfo {volume} {12}},\ \bibinfo {pages} {626793} (\bibinfo
  {year} {2021})}\BibitemShut {NoStop}%
\bibitem [{\citenamefont {Kaplinsky}\ and\ \citenamefont
  {Arnaout}(2016)}]{kaplinsky2016}%
  \BibitemOpen
  \bibfield  {author} {\bibinfo {author} {\bibfnamefont {J.}~\bibnamefont
  {Kaplinsky}}\ and\ \bibinfo {author} {\bibfnamefont {R.}~\bibnamefont
  {Arnaout}},\ }\bibfield  {title} {\bibinfo {title} {Robust estimates of
  overall immune-repertoire diversity from high-throughput measurements on
  samples},\ }\href {https://doi.org/10.1038/ncomms11881} {\bibfield  {journal}
  {\bibinfo  {journal} {Nature Communications}\ }\textbf {\bibinfo {volume}
  {7}},\ \bibinfo {pages} {11881} (\bibinfo {year} {2016})}\BibitemShut
  {NoStop}%
\end{thebibliography}%


\end{document}